\documentclass[11pt]{article}
\usepackage{latexsym,amssymb}
\usepackage{graphics}
\usepackage{pstricks}
\usepackage{amsmath}
\usepackage{subfigure}
\usepackage{vmargin}  
\usepackage{epsfig}
\usepackage[small,hang]{caption2}

 \setmarginsrb{2.cm}{2.cm}{2.cm}{2.5cm}{0cm}{0.cm}{0cm}{1.cm}

\begin{document}

\begin{flushright}
PCCF RI 0219\\
CPPM-C-2002-02\\
November 2002
\end{flushright}

\vspace{1.cm}

\begin{center}
{\Large {\bf Indirect Detection of Neutralino Dark Matter}}
\vspace{0.2cm}\\
{\Large {\bf with Neutrino Telescopes\footnote{``Mini review'' talk given at the 10th International Conference on Supersymmetry and Unification \\of Fondamental Interactions, DESY, Hambourg, Germany, June 17-23, 2002.}.}}
\vspace{0.3cm}\\
{\large E. Nezri}
\vspace{0.3cm}\\
 Centre de Physique des Particules de Marseille\\
IN2P3-CNRS, Universit\'e de la M\'editerran\'ee, F-13288 Marseille Cedex 09
\vspace{0.2cm}\\
 Laboratoire de Physique Corpusculaire de Clermont-Ferrand\\ 
IN2P3-CNRS, Universit\'e Blaise Pascal, F-63177 Aubiere Cedex
\vspace{0.2cm}\\
{\tt email : nezri@in2p3.fr}

\end{center}

\abstract{Principle of neutralino dark matter detection with neutrino telescopes and predictions are reviewed. The future Antares detector is described. Prospection in the CMSSM is exposed including comparison with experiment sensitivities of both neutrino indirect detection and direct detection experiments. Results in other frameworks are shortly surveyed. Models with a non negligible neutralino higgsino fraction are promising for neutrino telescopes detection.

\section{Neutralino Dark matter in neutrino telescopes}

Today, both theory and experimental data in cosmology suggest and focus on a $\Lambda CDM$ flat and black universe with the amounts of dark energy and cold dark matter: $\Omega_{\Lambda}\sim 0.7$, $\Omega_{CDM}\sim 0.3$. Dark matter would be a bath of Weakly Interacting (to fullfill a relic density of order 1) Massive Particles: WIMPs which are non relativistic since their decoupling. In the MSSM framework, assuming R-parity conservation, the Lightest Supersymmetric Particle (LSP) is stable and is the lightest neutralino ($\equiv$ {\it the} neutralino(s) $\chi$) in most regions of the parameter space. If present in galactic halos, relic neutralinos must accumulate in astrophysical bodies (of mass $M_b$) like the Sun or the Earth (see \cite{Jungman:1996df} for a review). The capture rate $C$ depends on the neutralino-quark elastic cross section:
$\sigma_{\chi-q}$. Neutralinos being Majorana particles, their vectorial
interaction vanishes and the allowed interactions are scalar (via $ \chi q
\xrightarrow{H,h} \chi q$ in $t$ channel and $ \chi q
\xrightarrow{\tilde{q}} \chi q$ in $s$ channel) and axial (via $\chi q
\xrightarrow{Z} \chi q$ in $t$ channel and $ \chi q \xrightarrow{\tilde{q}}
\chi q$ in $s$ channel). Depending on the spin content of the nuclei $N$
present in the body, scalar and/or axial interactions are
involved. Roughly, $C\sim \frac{\rho_{\chi}}{v_{\chi}} \sum_NM_bf_N
\frac{\sigma_N}{m_{\chi}m_N} <v^2_{esc}>_N
F(v_{\chi},v_{esc},m_{\chi},m_N)$, where $\rho_{\chi}, v_{\chi}$ are the
local neutralino density and velocity, $f_N$ is the density of nucleus $N$
in the body, $\sigma_N$ the nucleus-neutralino elastic cross section,
$v_{esc}$ the escape velocity and $F$ a suppression factor depending on
masses and velocity mismatching.  Considering that the population of captured neutralinos has a
velocity lower than the escape velocity, and therefore neglecting evaporation,
the number $N_{\chi}$ of neutralinos in the centre of a massive
astrophysical object depends on the balance between capture and
annihilation rates: $\dot{N_{\chi}}=C-C_AN_{\chi}^2$, where $C_A$ is the
total annihilation cross section $\sigma^A_{\chi-\chi}$ times the relative velocity per volume. The
annihilation rate at a given time $t$ is then:
\begin{equation}
\Gamma_A=\frac{1}{2}C_AN_{\chi}^2=\frac{C}{2}\tanh^2{\sqrt{CC_A}t}
\label{anihirate}
\end{equation}
with $\Gamma_A\approx \frac{C}{2}=cste$ when the neutralino population has
reached equilibrium, and $\Gamma_A\approx \frac{1}{2}C^2C_At^2$ in the
initial collection period (relevant in the Earth). So, when accretion is
efficient, the annihilation rate does not depend on annihilation processes but follows the capture rate $C$ and thus the neutralino-quark elastic cross section.\\The neutrino differential flux resulting of $\chi\chi$ annihilation is given by:
\begin{equation}
\frac{d\Phi}{dE}= \frac{\Gamma_A}{4 \pi R^2}\sum_FB_F\left(\frac{dN}{dE}\right)_{F}
\label{eq:nuflux}
\end{equation}
where $R$ is the distance between the source and the detector, $B_F$ is the branching ratio of annihilation chanel $F$ and $(dN/dE)_F$ its differential neutrino spectrum.
As $\chi\chi\rightarrow \nu\bar{\nu}$ is strongly suppressed by the tiny
neutrino mass, neutrino fluxes come from decays of primary annihilation
products, with a mean energy $E_{\nu}\sim\frac{m_{\chi}}{2}$ to
$\frac{m_{\chi}}{3}$. The most energetic spectra, called ``hard'' come from
neutralino annihilations into $WW$, $ZZ$ and the less energetic, ``soft'',
ones comes from $b\bar{b}$. Muon neutrinos give rise
to muon fluxes by charged-current interactions in the Earth. As both the
$\nu$ charged-current cross section on nuclei and the produced muon range
are proportional to $E_{\nu}$, high energy neutrinos are easier to
detect.

As we will compare neutrino telescope and direct detection experiment sensitivities, we can notice here that the elastic cross section of a neutralino on a nucleus also depends on $\sigma_{\chi-q}$, the nucleus mass number $A$ and its spin content. Depending on the chosen nuclei target, the detection rate of a direct detection experiment will be determined by those parameters.
\begin{figure}[t]
 \begin{center}
\includegraphics[width=0.8\textwidth]{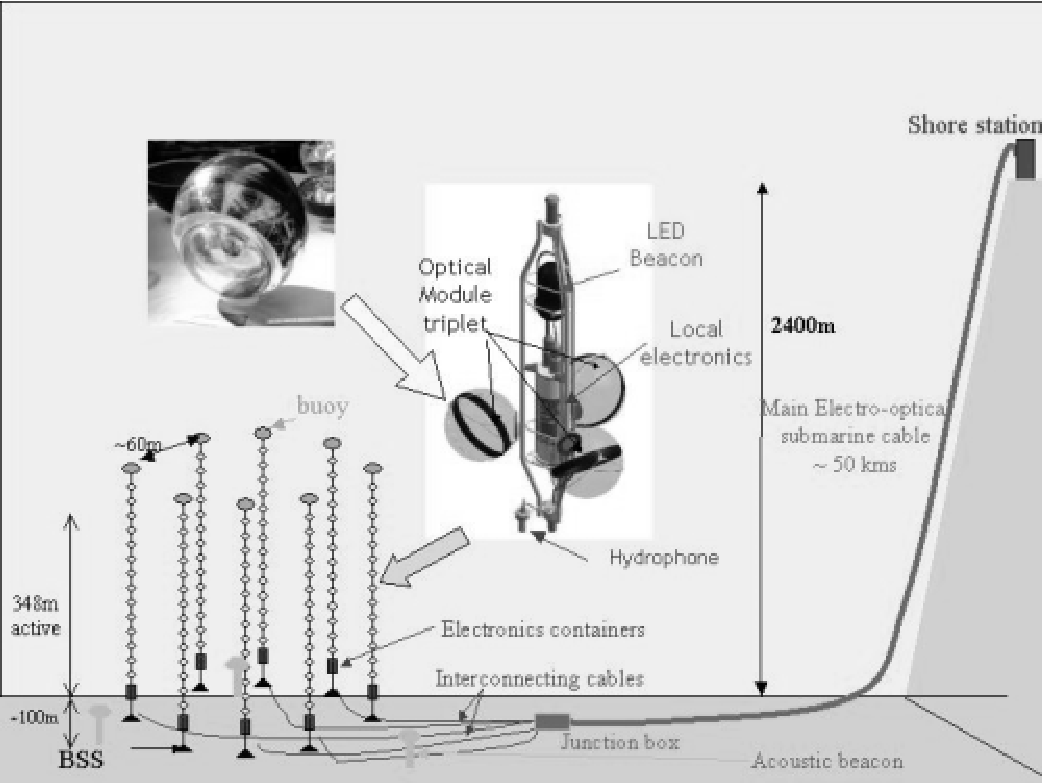}
 \caption{\small Schematic overview of the Antares detector showing the dimensions and separations of the elements  and the structure of the triplets of optical modules at each storey of the detector.}  
\label{antares}
 \end{center} 
\end{figure}
\subsection*{Neutrino telescope principle : the Antares detector}
Those kind of detection use the Earth as a target to convert neutrino fluxes into muons by charged current interactions. 
The muons produced emit \v{C}erenkov light when travelling through deep water or ice. Then the detection principle is to have a 3-dimensional array of photomultipliers collecting muon \v{C}erenkov light from which direction and energy of the muons(neutrinos) can be reconstructed. The international Antares collaboration is constructing such a 0.1 ${\rm km^2}$ telescope in Mediterranean sea at a depth of 2.4 km down south of Toulon (France). It will consits of 12 independent lines. Each line supports 90 optical modules which are water pressure resistant glass sphere including the photomultiplier tubes and the associated electronic material (see figure \ref{antares}). The Antares sensitivity to CDM neutralino annihilation in a ponctual source (Sun, Galactic Center) is obtained by requiring at least 2 lines hitted by \v{C}erenkov light. The atmospheric neutrino background is suppressed by looking in an optimised angular cone (2-3 degrees) in the source direction. The neutrino spectrum shape is calculated following \cite{Agrawal:1996gk} for the background (this gives around 1-2 events in the cone per year) and assuming a ``hard'' spectrum for the signal. Results on the Antares sensitivity on muon fluxes coming from neutralino annihilation coming from the Sun are shown on figure \ref{expvsmsug}a).

\section{Prospection in CMSSM/mSugra}
\begin{figure}[!h]
\begin{center}
 \includegraphics[width=0.9\textwidth]{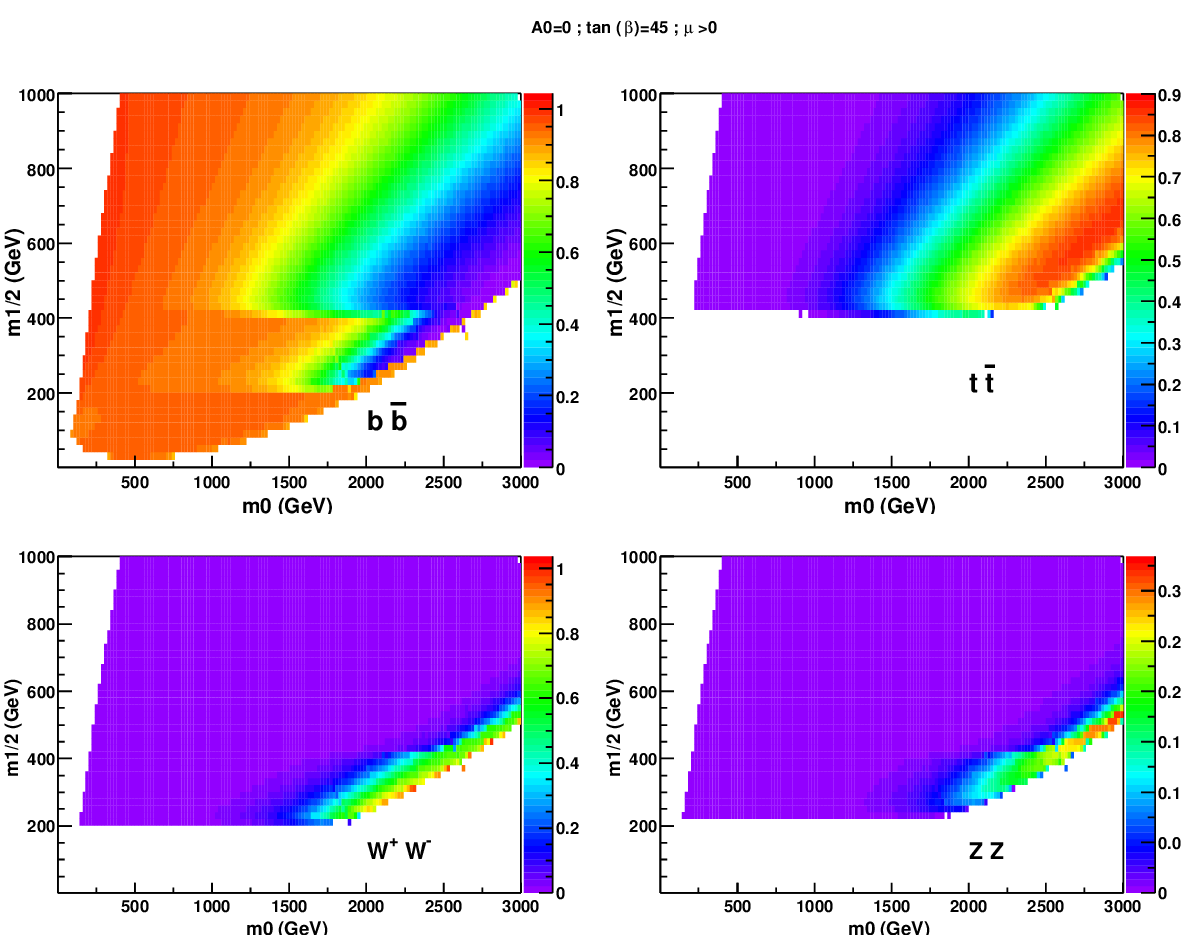}
 \caption{\small Dominant branching of the neutralino annihilation in the $(m_0,m_{1/2})$ plane.}  
\label{bratio}
\end{center}
\end{figure}
Recent studies in this framework can be found in \cite{Feng:2000gh,Barger:2001ur,Bertin:2002ky}.
Those models, inspired by gravity-mediated supersymmetry
breaking in supergravity \cite{Nilles:1984ge}, assume Grand
Unification of soft terms at $E_{GUT}\sim2.10^{16}$ GeV parameterized by
$m_0$ (common scalar mass), $m_{1/2}$ (common gaugino mass), $A_0$
(common trilinear term), $B_0$ (common bilinear term), and the Higgs mass parameter $\mu_0$. Requiring radiative electroweak symmetry breaking 
\begin{equation}\label{potminimi}
\frac{1}{2}m^2_Z=\frac{m^2_{H_d}\vert_{Q_{EWSB}}
	-m^2_{H_u}\vert_{Q_{EWSB}}\tan^2\beta}
	{\tan^2\beta-1}-\mu^2\vert_{Q_{EWSB}}
	{\underset{\tan{\beta}\gtrsim5}\sim}
	-m^2_{H_u}\vert_{Q_{EWSB}}-\mu^2\vert_{Q_{EWSB}}
	\ ,
\end{equation}
the usual input parameters of a constraint MSSM or mSugra model \cite{Chamseddine:1982jx,Barbieri:1982eh,Hall:1983iz} are
$$
m_0,\ m_{1/2},\ A_0,\ \tan\beta,\ sgn(\mu).
$$

In those models the neutralino can exhibit two different  natures, depending on the input parameter values : 
\begin{itemize}
\item[-] an almost pure bino-like neutralino for low $m_0$, as the RGE
drive \\$M_1\vert_{Q_{EWSB}} \simeq 0.41M_1\vert_{GUT} = 0.41m_{1/2} <<
|\mu|_{Q_{EWSB}}$ and \\$M_2\vert_{Q_{EWSB}} \simeq 0.83M_1\vert_{GUT} =
0.83m_{1/2} << |\mu|_{Q_{EWSB}}$;
\item[-] for $m_0\gtrsim1000$ GeV, the neutralino picks up some higgsino
mixing for $\tan{\beta}\gtrsim5$ as the increase of $m_0$ drives
$m^2_{H_u}$ to less negative values so that both $|m^2_{H_u}|$ and $|\mu|$
(via eq. \ref{potminimi}) decrease \cite{Feng:1999zg,Feng:2000gh}. One can then have $ |\mu|_{Q_{EWSB}}
\lesssim M_1\vert_{Q_{EWSB}} $ depending on $m_{1/2}$. When $|\mu|$ is too
small, EWSB cannot be achieved.
\end{itemize}
\subsection*{Neutralino annihilation}
In a $(m_0,\ m_{1/2})$ plane one has typically 4 dominant branching ratios for neutralino annihilation \cite{Bertin:2002ky}. A wide region of annihilation into $b\bar{b}$ when the neutralino is bino like and when the higgsino component is not negligible annihilation into gauge bosons $W^+W^-(ZZ)$ if $m_{\chi}<m_t$ or into $t\bar{t}$ if $m_{\chi}>m_t$. This is shown on figure \ref{bratio} for $A_0=0$ and $\tan{\beta}=45$. Varying their value only change size and shape of the 4 regions ({\it e.g} lower value of $\tan{\beta}$ increase the pseudo-scalar mass $m_A$ and reduce the $b\bar{b}$ region).
\subsection*{Capture and direct detection}
{\it For the Earth}, scalar interactions
dominate. By increasing $m_0$ from its low value region, sfermions and $H$
exchanges first decrease, and the scalar neutralino-proton cross section rises again when
approaching the mixed higgsino region (see figure \ref{sigpsi}a).

{\it For the Sun}, the spin of hydrogen allows for axial interaction, which
are stronger due to the $Z$ coupling. The latter depends strongly on the
neutralino higgsino fraction and is independent of $\tan{\beta}$, so the
spin dependent neutralino proton cross section follows the higgsino fraction isocurves along the ``no EWSB'' region (figure \ref{sigpsi}b).

Depending on the spin of the experiment nuclei target, direct detection rates follow scalar or spin dependent neutralino-proton cross section $\sigma^{scal/spin}_{\chi -p}$.
\begin{figure}[!h]
\begin{center}
 \begin{tabular}{cc}
 \includegraphics[width=0.46\textwidth]{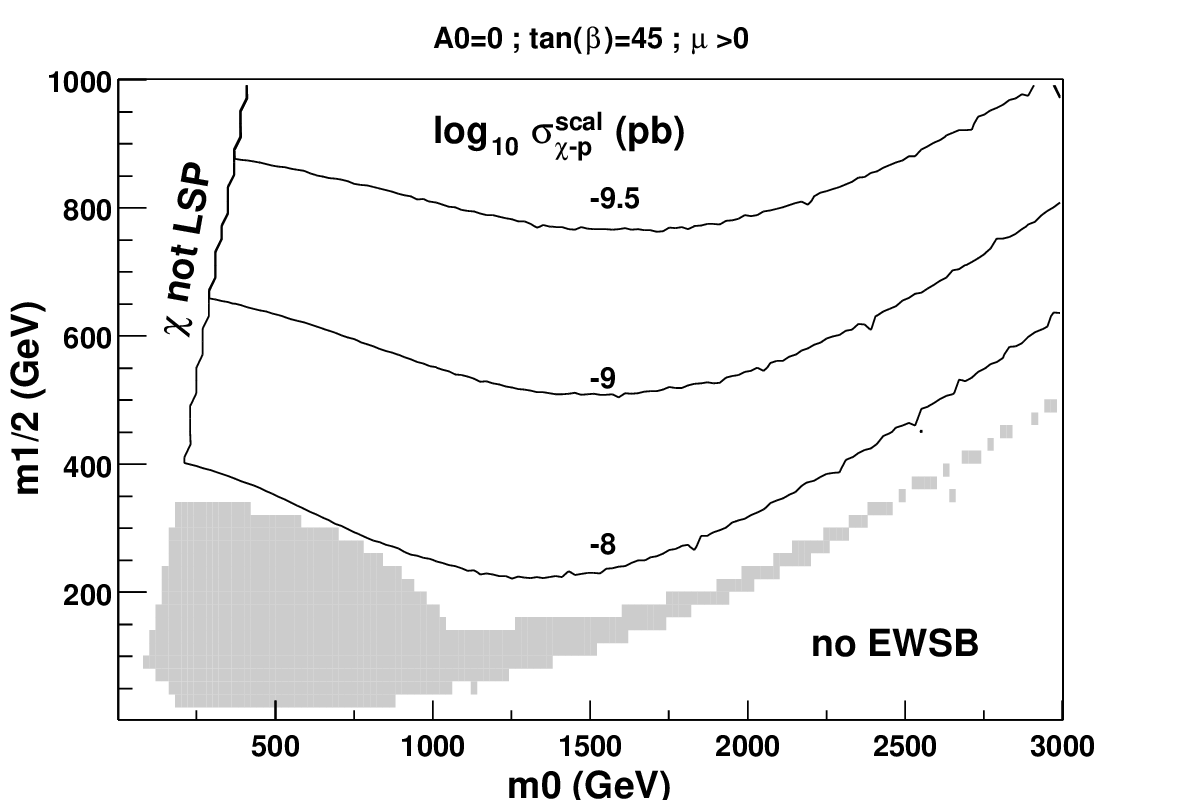} &
\includegraphics[width=0.46\textwidth]{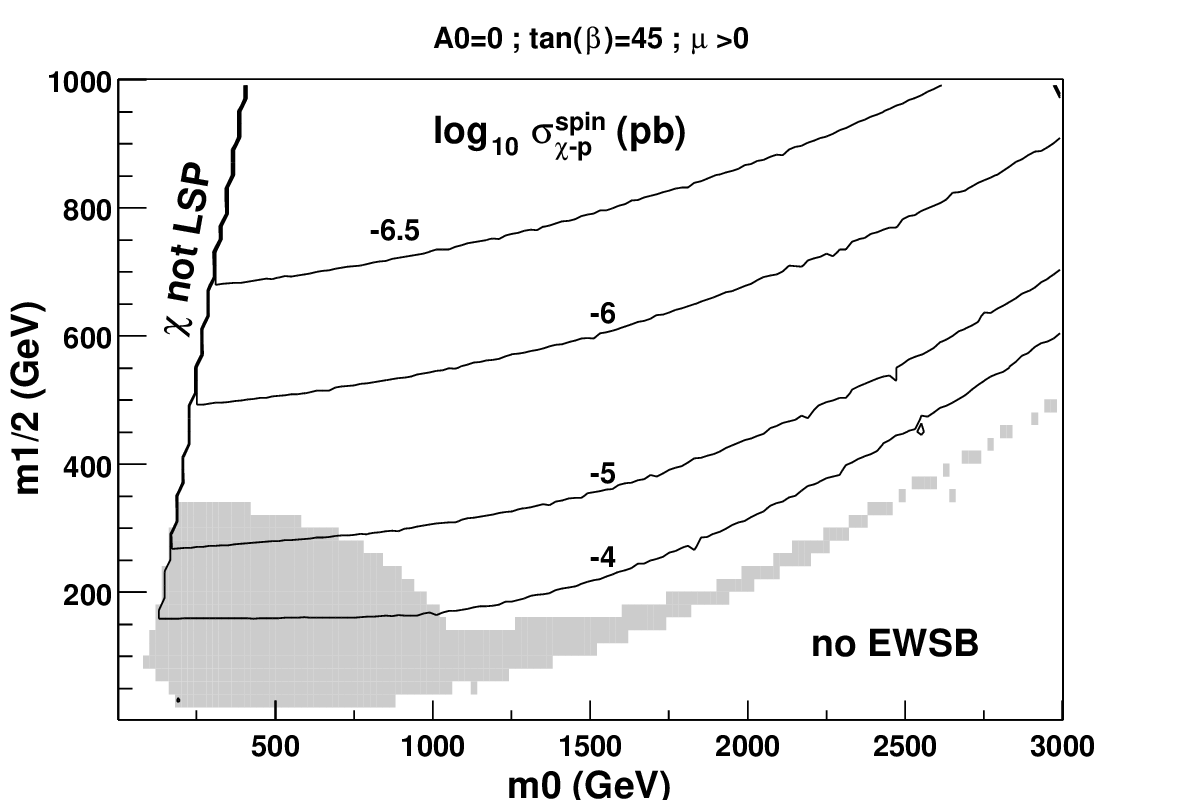}\\
a) & b)
\end{tabular}
 \caption{\small Scalar (a) and  axial (b) cross sections for neutralino
   scattering on proton in pb. Grey region is excluded as explicited in \cite{Bertin:2002ky} by experimental results (limits on SUSY masses, $m_h<114$ GeV, branching ratio $b\rightarrow s + \gamma\  \not\in\ [2.3\times10^{-4};5.3\times10^{-4}]$ and SUSY contribution to muon anomalous magnetic moment $a^{SUSY}_{\mu}\ \not\in\ [-6\times10^{-10};58\times10^{-10}]$) }
 \label{sigpsi}
 \end{center}
\end{figure}

\subsection*{Neutrino fluxes}

{\it For the Sun} (figure \ref{sunflux}), neutralinos do not reach complete equilibrium in the whole
$(m_0,m_{1/2})$ plane studied, actually when capture is low (see figure \ref{sigpsi}b) ).  
\begin{figure}[!h]
\begin{center}
\begin{tabular}{c}
 \includegraphics[width=\textwidth]{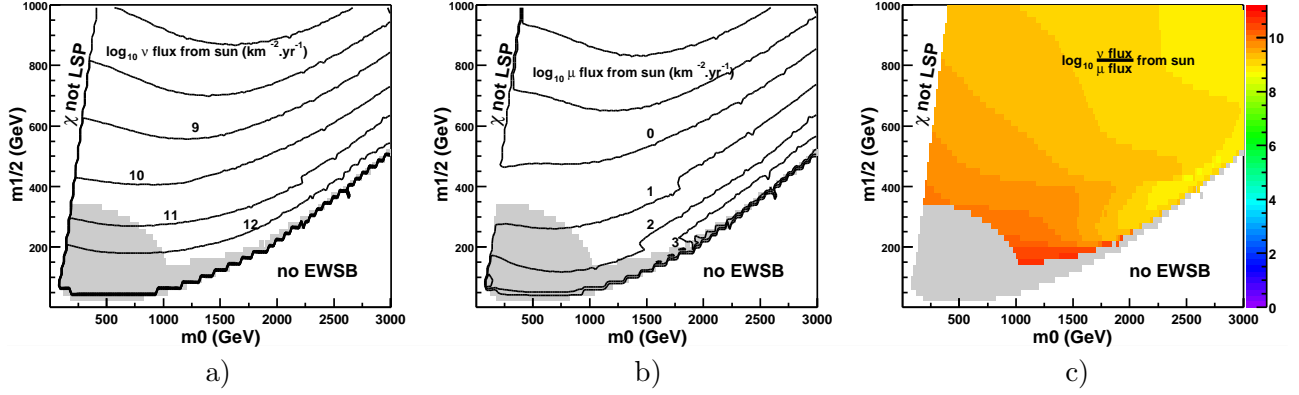}\\
  a) \hspace{0.3\textwidth} b) \hspace{0.3\textwidth} c)
 \end{tabular}
\caption{\small $\nu$ fluxes from the Sun (a), the corresponding $\mu$ fluxes with $E_{\mu}>5$ GeV theshold (b) and their ratio (c).}
 \label{sunflux}
 \end{center}
\end{figure}
For high $m_0$ or small $m_{1/2}<600$~GeV values, as equilibrium is
nevertheless reached, fluxes follow essentially the higgsino fraction and
the spin dependent $\sigma^{spin}_{\chi-p}$ isocurves of figure \ref{sigpsi}b). For low $m_0$ however, the equilibrium fluxes
would drop with $m_0$ and $C$, and $\Gamma_A$ feels the annihilation cross section behavior.
This effect is stronger for high $m_{1/2}$ values where the neutralinos are
heavier and more bino-like, and where capture is smaller. The effect is also
larger when $\tan{\beta}$ (and thus $\sigma^A_{\chi-\chi}$) is low, as
incomplete equilibrium makes $\nu(\mu)$ fluxes sensitive to
$\sigma^A_{\chi-\chi}$, making the $W$ and top thresholds more conspicuous.

{\it For the Earth} (figure \ref{earthflux}), neutralinos are not in equilibrium. Neutrino fluxes
depend both on $C^2$ and annihilation, giving an enhancement in the low and
high $m_0$ regions where fluxes are boosted by annihilation (see figure
\ref{sigpsi} and \ref{earthflux}). Since $M_{\bigoplus}<M_{\bigodot}$ and
$\sigma^{scal}_{\chi-p}<\sigma^{spin}_{\chi-p}$, the capture rate and $\nu$
fluxes from the Earth are much smaller than from the Sun.\\

\begin{figure}[!h]
\begin{center}
\begin{tabular}{c}
 \includegraphics[width=\textwidth]{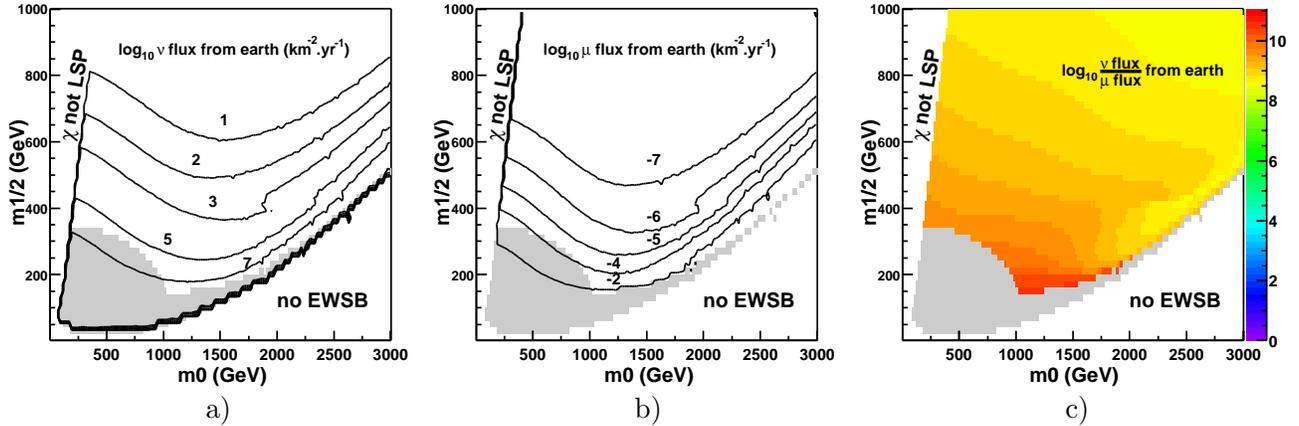}\\
  a) \hspace{0.3\textwidth} b) \hspace{0.3\textwidth} c)
 \end{tabular}
\caption{\small $\nu$ fluxes from the Earth (a), the corresponding $\mu$
fluxes $E_{\mu}>5$ GeV theshold (b) and their ratio (c).}
\label{earthflux} \end{center}
\end{figure}

{\it Comparing $\nu$ fluxes and $\mu$ fluxes}, the $\nu\to\mu$ conversion
factor increases with $m_{1/2}$ due to the increase of $m_{\chi}$ leading
to more energetic neutrinos. This ratio also follows the annihilation final
state regions described above. Indeed, spectra from $WW$, $ZZ$
and $t\bar{t}$ are more energetic than $b\bar{b}$
spectra, so neutrino conversion into muon is more efficient in the mixed
bino-higgsino region above $W$ and top thresholds (see figure \ref{sunflux}
and \ref{earthflux}).\\

\subsection*{Comparison with Experimental sensitivities}
Because of the very low value of $\nu(\mu)$ fluxes coming from neutralino annihilation in the Earth (figure \ref{earthflux}), current and next generation experiments are not able to test the CMSSM on this signal. We can only compare current experiment sensitivities with the predictions for the Sun.
\begin{figure}[h!]
\begin{center}
\begin{tabular}{c}
 \includegraphics[width=0.9\textwidth]{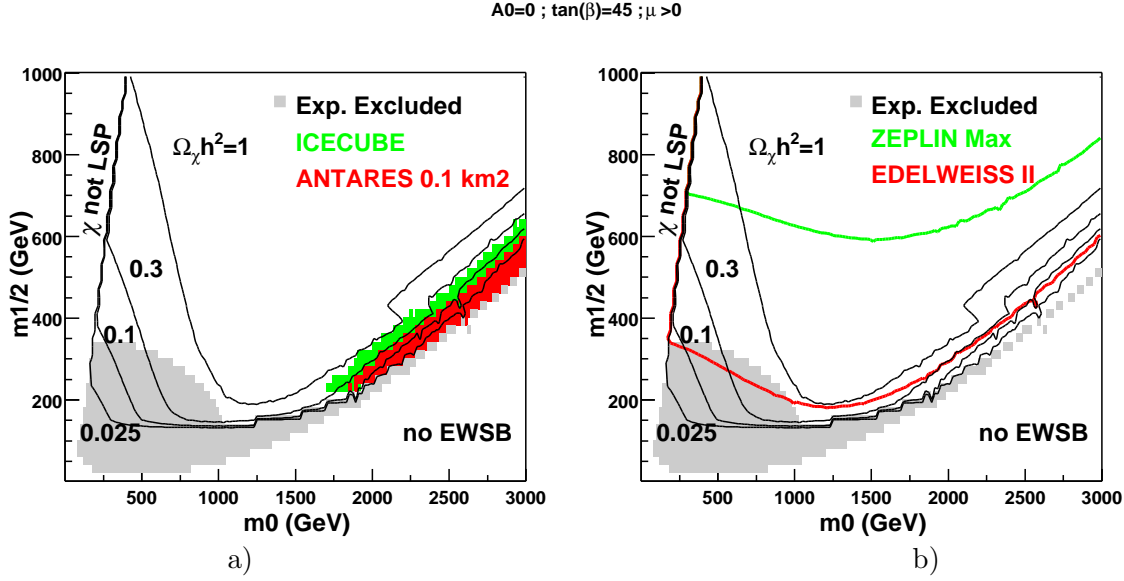}\\ a)
  \hspace{0.5\textwidth} b) \end{tabular} \caption{\small a) $\nu$
  telescopes sensitivities on $\mu$ fluxes from the Sun and b) direct
  detection experiments sensitivities in the $(m_0,m_{1/2})$ plane.}
\label{manip}
   \end{center}
\end{figure}
\begin{figure}[!h]
\begin{center}
 \begin{tabular}{cc}
 \includegraphics[width=0.45\textwidth]{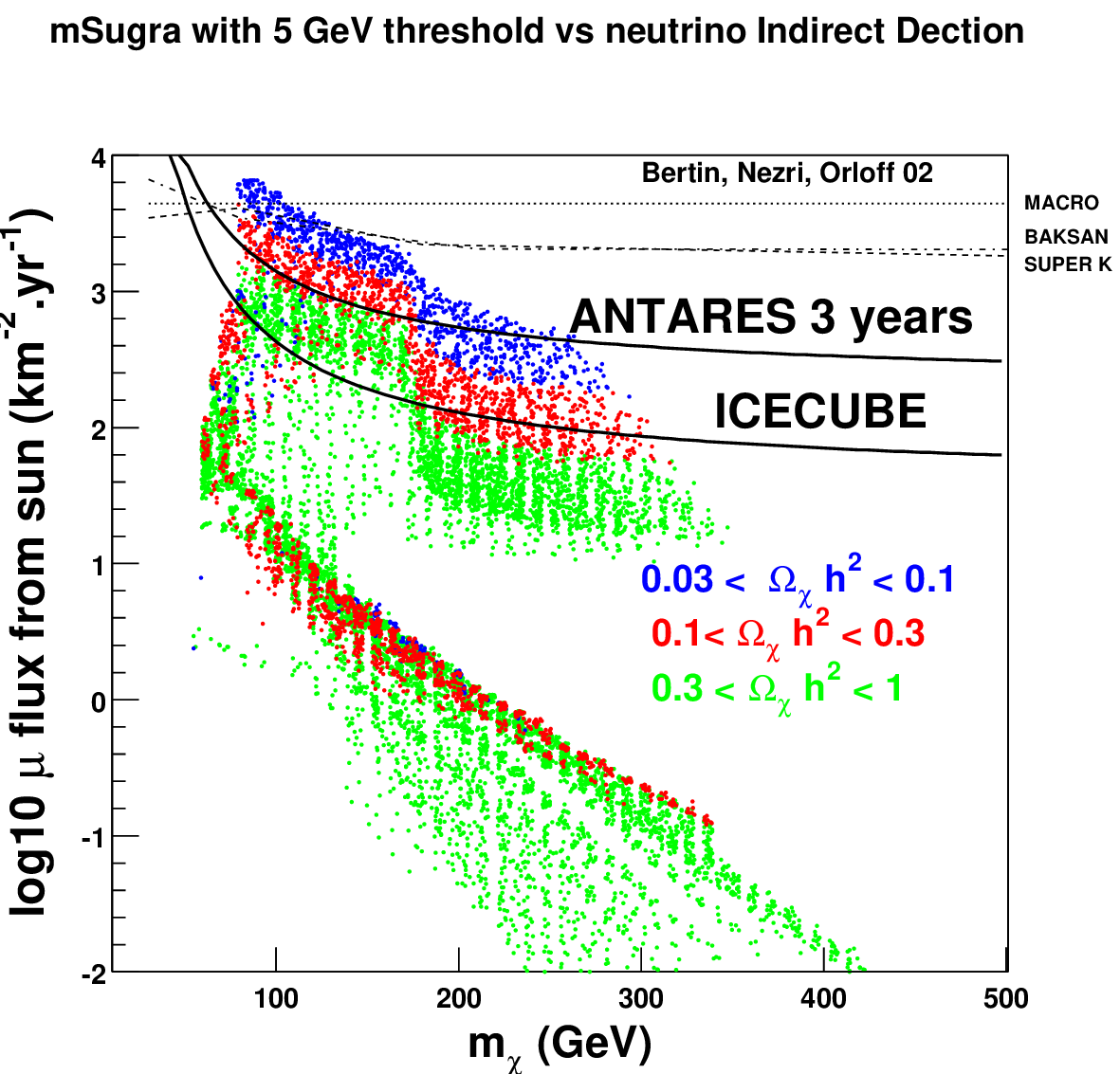} &
\includegraphics[width=0.45\textwidth]{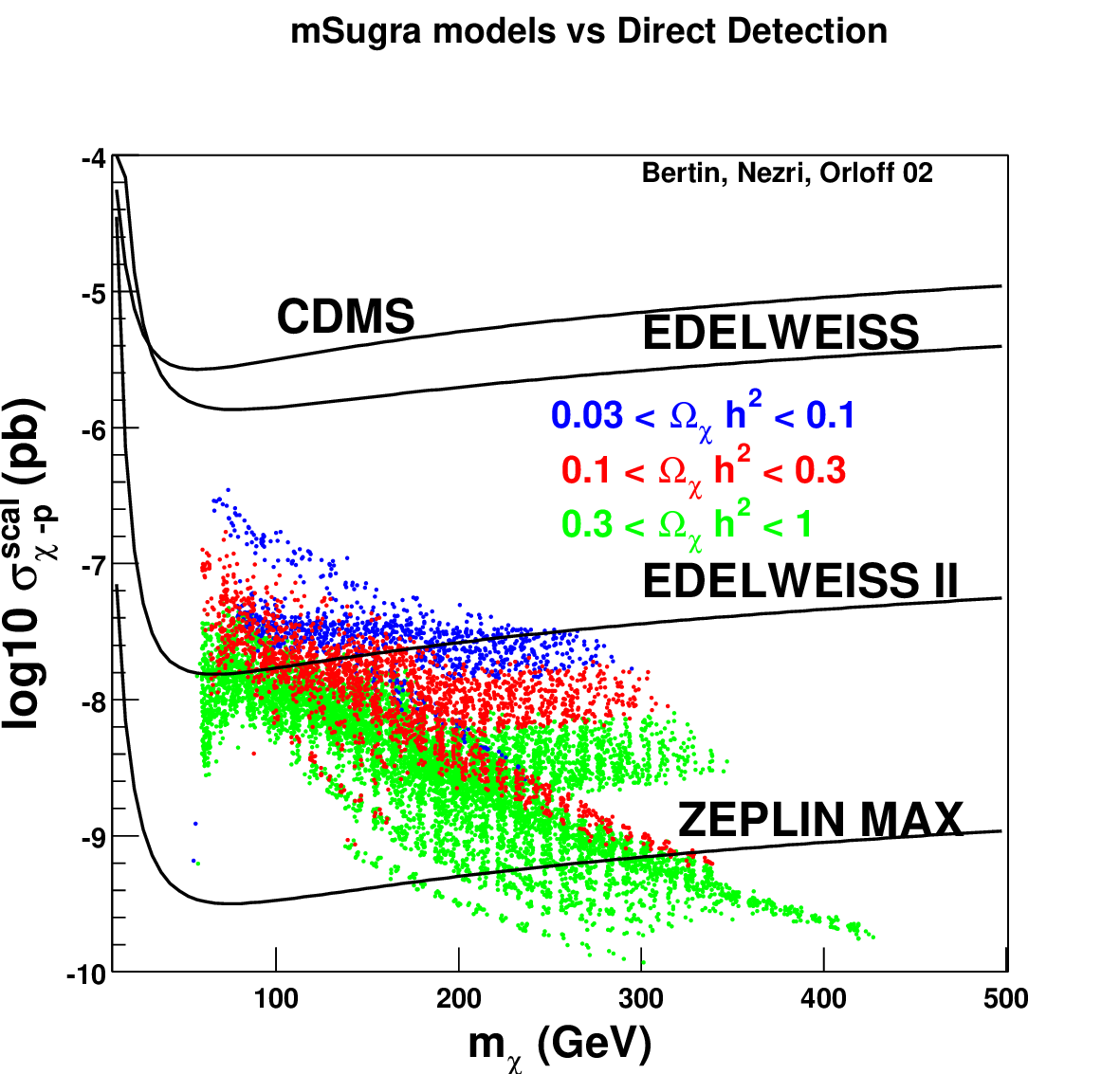}\\
a) & b)
\end{tabular}
 \caption{\small Neutrino indirect detection a) and direct detection b) sensitivities in the CMSSM.}
 \label{expvsmsug}
 \end{center}
\end{figure}

Sensitivities of both neutrino indirect detection and direct detection experiments in the $(m_0,m_{1/2})$ plane are shown on figure \ref{manip} for $\tan{\beta}=45$, $A_0=0$ and in the $({\rm signal} ; m_{\chi})$ plane on figure \ref{expvsmsug} for a wide sample of models. MSugra models predicting a good relic density of
   neutralinos give neutrino/muon fluxes which can be as high as the current
   experimental limits (Baksan \cite{Suvorova:1999my}, Macro \cite{Macro},
   Super Kamiokande \cite{SuperK}). The $0.1 \ {\rm km^2}$ Antares detector
   will explore further the interesting parameter space \cite{AntarLee}. Next
   generation neutrinos telescopes (Icecube, Antares ${\rm km^3}$) will be
   much more efficient to test such models, {\it e.g} Icecube expected sensitivity
   $\sim10^2\ \mu\ {\rm km^2 \ yr^{-1}}$ from the Sun \cite{Ice3Edsjo}.\\

{\it Indirect vs Direct detection} (figures \ref{manip} and \ref{expvsmsug}): a high neutralino-proton cross section
is efficient in both direct and indirect detection (via capture). The large
$m_0$ mixed higgsino-bino region pointed out in
\cite{Feng:2000gh,Feng:2000zu} favours both direct and indirect detection
due to the enhancement of $\sigma_{\chi-p}$ and
$\sigma^A_{\chi-\chi}$. Both enter in the indirect detection which is
moreover favoured by the production of more energetic neutrinos in $WW,\
ZZ$ and $t\bar{t}$ decays, leading to better conversion into muons. This mixed region, which has a good relic
density, is very attractive for neutrino indirect detection signal.

Hovewer, no current experiment is able to really test such models.

\section{Short survey of other frameworks}
\subsection*{Low energy MSSM}
Severals studies \cite{Jungman:1996df,Bergstrom:1998xh,Barger:2001ur} have been done in a low energy MSSM approach with more free parameters than in mSugra/CMSSM, especially $m_A$ and $\mu$, the latter leading more easily to an higgsino fraction as a dominant effect and high muon fluxes. This is shown on figure \ref{mssmkao} extracted from \cite{Barger:2001ur}. It is clear that this kind of models exhibit less consistency and theoretical motivations but they allow to explore all phenomenologies that can not be tested with high energy models because of the RGE evolution giving rise to a quite hierarchic SUSY spectrum.

\begin{figure}[!h]
 \begin{center} 
 \includegraphics[width=0.7\textwidth]{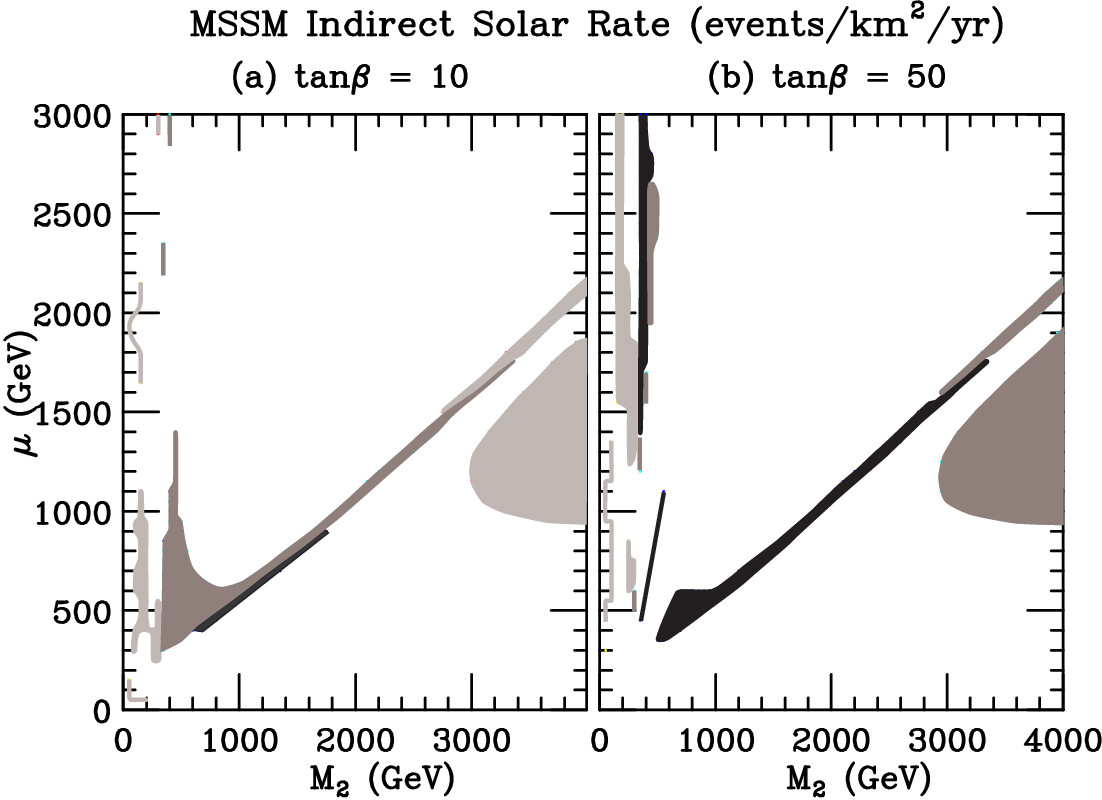}
 \caption{\small Low energy MSSM \cite{Barger:2001ur}: Regions of indirect detection rate
of the neutralino dark matter in the ($M_2,\mu$) plane
with $M_{\rm SUSY} = {\rm MAX}(300 \; {\rm GeV}, 1.5 m_{\chi^0_1})$
for (a) $\tan\beta = 10$ and (b) $\tan\beta = 50$.
The region with dark shading has $dN_{\rm ID}/dA > 100$;
the region with light shading has $dN_{\rm ID}/dA < 10$; and
the region with intermediate shading has $100 > dN_{\rm ID}/dA > 10$
(event/km$^2$/year).
The blank regions do not have a cosmologically interesting relic density
($0.05 \leq \Omega_{\chi^0_1} \leq 0.3$) for the neutralino dark matter.}  
\label{mssmkao}
 \end{center} 
\end{figure}

\subsection*{Non universality at GUT scale}
An other possibility is to relax universality at high energy, keeping RGE traitement and radiative electroweak symmetry breaking \cite{Bertin:2002sq}. 

This effect in the Higgs sector ($m^2_{H_1}|_{GUT}\not=m^2_{H_2}|_{GUT}$) has been studied in \cite{Berezinsky:1996ga,Barger:2001ur,Ellis:2002iu,Bertin:2002sq} and allows lower values of $\mu$ and so higher muon fluxes more easily than in the universal case, but in a localised region. Typically in a $(m_0,m_{1/2})$ plane the region where radiative electroweak symmetry breaking {\it can not} be achieved is wider and  the gradient of $\mu$ values along the boundary is more steeper than in mSugra/CMSSM. 

Changing high energy gaugino mass relations can lead to more interesting effect, especially lowering $M_3|_{GUT}$ lead to wider region with cosmologically favoured neutralino relic density with strong enhancement of both direct and indirect detection rates \cite{Bertin:2002sq}.

\section{Conclusion}
Muon fluxes resulting from neutralino annihilations in the centre of the Earth are typically small. In RGE models, they are too low to be detected because of heavy scalars lowering capture rate ($\sigma^{scal}_{\chi-p}$), but the fluxes from this source can be enhanced in low energy models. The Sun case can be interesting when the neutralino higgsino fraction lead to both a good relic density and high detection rates. This happends for heavy scalars along the radiative electroweak symmetry breaking boundary of the $(m_0,m_{1/2})$ plane in RGE models. Current neutrino indirect detection experiments are not sensitive enough to constrain RGE models, but futur projects will be very efficient to test the Sun signal. This kind of information will be complementarity with accelerators since TeV's scalars are beyond reach of Tevatron or LHC. One also has to notice that RGE models giving high fluxes from the Sun have very small contribution to muon anomalous magnetic moment $g(-2)_{\mu}$ and would be ruled out in case of a real exclusion of $(g-2)^{SM}_{\mu}$. To finish a remark should be adressed on different predictions between authors mainly due to differences on the crucial $\mu$ parameter values due to differences between RGE codes in potential minimization \cite{Allanach:2002pz}.\\

{\bf Acknowledgement}\\
I thank the organisers for invitation to give this mini review talk including the Antares collaboration, C. Kao {\it et al} \cite{Barger:2001ur} and \cite{Bertin:2002ky} contributions. I also thank my supervisors Vincent Bertin and Jean Orloff.

\nocite{}
\bibliography{nezri_SUSY02proc}

\providecommand{\href}[2]{#2}\begingroup\raggedright\begin{thebibliography}{10}

\bibitem{Jungman:1996df}
G.~Jungman, M.~Kamionkowski, and K.~Griest, {\it Supersymmetric dark matter},
  {\bf Phys. Rept.} {\bf 267} (1996) 195--373,
  [\href{http://xxx.lanl.gov/abs/http://arXiv.org/abs/hep-ph/9506380}{{\tt
  hep-ph/9506380}}].

\bibitem{Agrawal:1996gk}
V.~Agrawal, T.~K. Gaisser, P.~Lipari, and T.~Stanev, {\it Atmospheric neutrino
  flux above 1 GeV},  {\bf Phys. Rev.} {\bf D53} (1996) 1314--1323,
  [\href{http://xxx.lanl.gov/abs/http://arXiv.org/abs/hep-ph/9509423}{{\tt
  hep-ph/9509423}}].

\bibitem{Feng:2000gh}
J.~L. Feng, K.~T. Matchev, and F.~Wilczek, {\it Neutralino dark matter in focus
  point supersymmetry},  {\bf Phys. Lett.} {\bf B482} (2000) 388--399,
  [\href{http://xxx.lanl.gov/abs/http://arXiv.org/abs/hep-ph/0004043}{{\tt
  hep-ph/0004043}}].

\bibitem{Barger:2001ur}
V.~D. Barger, F.~Halzen, D.~Hooper, and C.~Kao, {\it Indirect search for
  neutralino dark matter with high energy neutrinos},  {\bf Phys. Rev.} {\bf
  D65} (2002) 075022,
  [\href{http://xxx.lanl.gov/abs/http://arXiv.org/abs/hep-ph/0105182}{{\tt
  hep-ph/0105182}}].

\bibitem{Bertin:2002ky}
V.~Bertin, E.~Nezri, and J.~Orloff, {\it Neutrino indirect detection of
  neutralino dark matter in the CMSSM},
  \href{http://xxx.lanl.gov/abs/http://arXiv.org/abs/hep-ph/0204135}{{\tt
  hep-ph/0204135}}.

\bibitem{Nilles:1984ge}
H.~P. Nilles, {\it Supersymmetry, supergravity and particle physics},  {\bf
  Phys. Rept.} {\bf 110} (1984) 1.

\bibitem{Chamseddine:1982jx}
A.~H. Chamseddine, R.~Arnowitt, and P.~Nath, {\it Locally supersymmetric Grand
  Unification},  {\bf Phys. Rev. Lett.} {\bf 49} (1982) 970.

\bibitem{Barbieri:1982eh}
R.~Barbieri, S.~Ferrara, and C.~A. Savoy, {\it Gauge models with spontaneously
  broken local supersymmetry},  {\bf Phys. Lett.} {\bf B119} (1982) 343.

\bibitem{Hall:1983iz}
L.~J. Hall, J.~Lykken, and S.~Weinberg, {\it Supergravity as the messenger of
  supersymmetry breaking},  {\bf Phys. Rev.} {\bf D27} (1983) 2359--2378.

\bibitem{Feng:1999zg}
J.~L. Feng, K.~T. Matchev, and T.~Moroi, {\it Focus points and naturalness in
  supersymmetry},  {\bf Phys. Rev.} {\bf D61} (2000) 075005,
  [\href{http://xxx.lanl.gov/abs/http://arXiv.org/abs/hep-ph/9909334}{{\tt
  hep-ph/9909334}}].

\bibitem{Suvorova:1999my}
O.~V. Suvorova, {\it Status and perspectives of indirect search for dark
  matter, Published in Tegernsee 1999, Beyond the desert 1999},
  \href{http://xxx.lanl.gov/abs/http://arXiv.org/abs/hep-ph/9911415}{{\tt
  hep-ph/9911415}}.

\bibitem{Macro}
{\bf MACRO} Collaboration, T.~Montaruli, {\it Search for WIMPs using
  upward-going muons in MACRO, Proceeedings of the 26th ICRC in Salt Lake City,
  hep-ex/9905021},
  \href{http://xxx.lanl.gov/abs/http://arXiv.org/abs/hep-ex/9905021}{{\tt
  hep-ex/9905021}}.

\bibitem{SuperK}
{\bf Super-Kamiokande} Collaboration, A.~Habig, {\it An indirect search for
  WIMPs with Super-Kamiokande},
  \href{http://xxx.lanl.gov/abs/http://arXiv.org/abs/hep-ex/0106024}{{\tt
  hep-ex/0106024}}.

\bibitem{AntarLee}
L.~Thompson, {\it Dark Matter Prospects with the ANTARES Neutrino Telescope,
  talk given at the conference DARK 2002, Cape Town, South Africa 4-9 Feb}, .

\bibitem{Ice3Edsjo}
J.~Edsjo, {\it Swedish Astroparticle Physics, talk given at the conference
  'Partikeldagarna', Uppsala, Sweden, March 6, 2001}, .

\bibitem{Feng:2000zu}
J.~L. Feng, K.~T. Matchev, and F.~Wilczek, {\it Prospects for indirect
  detection of neutralino dark matter},  {\bf Phys. Rev.} {\bf D63} (2001)
  045024,
  [\href{http://xxx.lanl.gov/abs/http://arXiv.org/abs/astro-ph/0008115}{{\tt
  astro-ph/0008115}}].

\bibitem{Bergstrom:1998xh}
L.~Bergstrom, J.~Edsjo, and P.~Gondolo, {\it Indirect detection of dark matter
  in km-size neutrino telescopes},  {\bf Phys. Rev.} {\bf D58} (1998) 103519,
  [\href{http://xxx.lanl.gov/abs/http://arXiv.org/abs/hep-ph/9806293}{{\tt
  hep-ph/9806293}}].

\bibitem{Bertin:2002sq}
V.~Bertin, E.~Nezri, and J.~Orloff, {\it Neutralino dark matter beyond CMSSM
  universality},
  \href{http://xxx.lanl.gov/abs/http://arXiv.org/abs/hep-ph/0210034}{{\tt
  hep-ph/0210034}}.

\bibitem{Berezinsky:1996ga}
V.~Berezinsky, A.~Bottino, J.~Ellis, N.~Fornengo, G.~Mignola, and S.~Scopel,
  {\it Searching for relic neutralinos using neutrino telescopes},  {\bf
  Astropart. Phys.} {\bf 5} (1996) 333--352,
  [\href{http://xxx.lanl.gov/abs/http://arXiv.org/abs/hep-ph/9603342}{{\tt
  hep-ph/9603342}}].

\bibitem{Ellis:2002iu}
J.~Ellis, T.~Falk, K.~A. Olive, and Y.~Santoso, {\it Exploration of the MSSM
  with non-universal Higgs masses},
  \href{http://xxx.lanl.gov/abs/http://arXiv.org/abs/hep-ph/0210205}{{\tt
  hep-ph/0210205}}.

\bibitem{Allanach:2002pz}
B.~Allanach, S.~Kraml, and W.~Porod, {\it Comparison of SUSY mass spectrum
  calculations},
  \href{http://xxx.lanl.gov/abs/http://arXiv.org/abs/hep-ph/0207314}{{\tt
  hep-ph/0207314}}.

\end{thebibliography}\endgroup
\bibliographystyle{manustyle}

\end{document}